\documentclass[reprint, superscriptaddress, groupedaddress, amsmath,amssymb, aps, prb]{revtex4-2}

\pdfoutput=1

\usepackage{graphicx}
\usepackage{bm}
\usepackage{hyperref}
\usepackage{xcolor}
\usepackage{url}
\usepackage[utf8]{inputenc}

\newcommand{\veck}{\mathbf{k}}
\newcommand{\ntop}{n_{\text{topo }}}
\newcommand{\ra}{\rightarrow}
\newcommand{\xik}{\xi_{\mathbf{k}}}
\newcommand{\diff}{\ensuremath{\mathrm{d}}}
\newcommand{\sk}{S_{\mathbf{k}}}
\newcommand{\kst}{\mathbf{k}^\ast}
\newcommand{\gk}{\mathcal{G}_{\mathbf{k}}}
\newcommand{\kc}{\mathbf{k}_c}

\begin{document}

\title{Quench dynamics and scaling laws in topological nodal loop semimetals}

\date{\today}

\author{Karin Sim}
\affiliation{Institute for Theoretical Physics, ETH Z\"{u}rich, 8093 Z\"{u}rich, Switzerland}
\author{R. Chitra}
\affiliation{Institute for Theoretical Physics, ETH Z\"{u}rich, 8093 Z\"{u}rich, Switzerland}
\author{Paolo Molignini}
\affiliation{T.C.M. group, Cavendish Laboratory, University of Cambridge, 19 J J Thomson Avenue, Cambridge CB3 0HE, United Kingdom}

\begin{abstract}
We employ quench dynamics as an effective tool to probe different universality classes of topological phase transitions. Specifically, we study a model encompassing both Dirac-like and nodal loop criticalities. 
Examining the Kibble-Zurek scaling of topological defect density, we discover that the scaling exponent is reduced in the presence of extended nodal loop gap closures.
For a quench through a  multicritical point, we also unveil a path-dependent crossover between two sets of critical exponents.
Bloch state tomography finally reveals additional differences in the defect trajectories for sudden quenches. 
While the Dirac transition permits a static trajectory under specific initial conditions, we find that the underlying nodal loop leads to complex time-dependent trajectories in general. 
In the presence of a nodal loop, we generically find a mismatch between the momentum modes where topological defects are generated and where dynamical quantum phase transitions occur. 
We also find notable exceptions where this correspondence breaks down completely. 

\end{abstract}

\maketitle

\section{Introduction}
Topological phase transitions (TPTs) refer to the changes in the global structure of ground state electron wavefunctions \cite{ti, ti2, ti4} accompanying a gap closure in the energy spectrum.  
The change in topological order across TPTs is signalled by a discrete jump of the corresponding topological invariant.
Depending on the dimensionality and symmetries, such invariants can often be obtained from the momentum space integration of an appropriate curvature function that encodes the curvature of a many-body state \cite{wei1}.
Well known examples include the Berry connection and Berry curvature \cite{berry}.    
Typically, the curvature function shows a universal divergence near the transition \cite{wei1, wei2}, which can be exploited to characterize the TPTs through the curvature renormalization group (CRG) method \cite{wei1, wei2, crg, crg2}. 
This renormalization group (RG)-like method characterizes the behavior of the curvature function near the critical point, which enables us to translate the notions of universality and critical exponents, seen in the context of standard symmetry-breaking phase transitions, to TPTs \cite{crg}. 
CRG showed the existence of different universality classes of TPTs in a broad range of topological insulators and superconductors, including static noninteracting systems, weakly and strongly interacting systems, and periodically driven systems \cite{crg}.

While methods such as CRG  provide a theoretical framework to devise critical exponents, an important question concerns the putative experimental relevance of these exponents.
In the case of standard phase transitions, quantum quenches \cite{quench} have been extensively used as a way of probing aspects of criticality \cite{polkonikov, Jafariquench, Jafariquench2}. 
Quenches are readily realizable in engineered systems such as ultracold atoms \cite{ucquench, becquench, latticequench} and superconducting circuits \cite{scquench}. 
In particular, they permit the visualization of defect generation in momentum space through time and momentum-resolved Bloch state tomography \cite{defectsexp1, defectsexp2, defectsexp3}. 

An important aspect of quenches across a symmetry-breaking phase transition is the Kibble-Zurek (KZ) mechanism \cite{kzm}, which predicts a simple scaling relation between the density of topological defects, the quench rate, and the underlying critical exponents characterizing the phase transition. 
More recently, sudden quenches  \cite{defectspra} which traverse a gap closure point in the phase diagram were also shown to generate dynamical quantum phase transitions (DQPTs) \cite{dqpt} in quantum many-body systems. DQPTs are the nonequilibrium counterpart of equilibrium phase transition and can be characterized by the nonanalyticity of the Loschmidt echo, which quantifies the overlap between the time-evolved and the initial wavefunctions, at critical times \cite{dqpt, dqpt3, Jafaridqpt, Jafaridqpt2}.

The quench dynamics and KZ scaling for TPTs with Dirac-like transitions have been well investigated both theoretically and experimentally \cite{dziarmaga,kyang, cooperquench, defectsexp1, defectsexp2, defectsexp3}.
In this case, there is a direct correspondence between the CRG and the KZ scaling exponents.
Defect dynamics and DQPTs across Dirac transitions, where defects are phase vortices in the Brillouin zone, have also been studied theoretically \cite{defectspra} and experimentally \cite{defectsexp1, defectsexp2, defectsexp3}.
On the other hand, not much work has been done on more exotic topological states such as nodal loop semimetals, which involve extended gap closures.
Nodal loop semimetals host a zoo of possible exotic properties, such as drumhead surface states and momentum-dependent transport properties \cite{nl1, nl2, nl3}. 
Nodal loop gap closures can also appear dynamically in periodically driven systems \cite{nlfloquet}, and can coexist with Dirac-type gap closures to generate exotic multicriticality, whose defect production might then be impacted by the presence of extended gap closure. 

In this paper, we investigate quench dynamics in a model hosting TPTs characterized by both Dirac and nodal loop gap closures, as well as multicriticality.  
We study the influence of the nature of the TPT on both the KZ-like scaling of topological defects and DQPTs. For the former, we uncover a link between the CRG and KZ scaling exponents, both of which depend on the functional dependence of the gap closure which defines the low-energy theory. In particular, we show that KZ scaling is dramatically altered by the existence of extended gap closures.
For the latter, we reveal a sensitive dependence of DQPTs and topological defects on the initial condition.

The paper is organized as follows: in Section \ref{sec:model}, we introduce our model Hamiltonian and its topological phase diagram.
In Section \ref{sec:methods}, we outline the theoretical concepts used to quantify the defect density and dynamics.
In Sections \ref{sec:dirac} and \ref{sec:nl}, we analyze Dirac and nodal loop transitions from the aspects of KZ scaling, defect production and dynamics, and their connections to DQPTs. 
We then extend our analysis to multicriticality in Section \ref{sec:mcp}, where both types of transitions are present.
Our conclusions and an outlook to future studies are presented in Section \ref{sec:outlook}.


\section{Model}
\label{sec:model}
We consider a minimal model for a semimetal containing both Dirac and nodal loop topological phase transitions (TPTs). 
The model consists of free fermions on a 2D square lattice with periodic boundary conditions (PBCs), described by the following Bogoliubov-de Gennes (BdG) Hamiltonian \cite{ham1, ham2, paolothesis, bogo, quantumising, dziarmaga}
\begin{equation}
 H_{\mathbf{k}} =
 	\begin{pmatrix}
        		\mu-\xi_{\mathbf{k}} & \lambda S_{\mathbf{k}} \\
        		\lambda S^\ast_{\mathbf{k}} & -(\mu-\xi_{\mathbf{k}})
        \end{pmatrix}
        \label{eqn:ham}
\end{equation}
where
\begin{equation}
	\begin{split}
		\xik &=2(\cos k_x+\cos k_y)    \\
		\sk &= \sin k_x+i\sin k_y.	
	\end{split}
	\label{eqn:sxi}
 \end{equation}
The Hamiltonian features a spectrum given by
\begin{equation}
	E_{\pm,\mathbf{k}}=\pm\sqrt{(\mu-\xik)^2+\lambda^2|\sk|^2}.
        	\label{eqn:disp}
\end{equation} 
We can also express Eqn (\ref{eqn:ham}) in the form
\begin{equation}
    	H_{\veck}=\mathbf{h_{\veck}}\cdot\mathbf{\sigma}
\end{equation}
where \(\sigma=(\sigma_x,\sigma_y,\sigma_z)^T\) is the Pauli  vector and
\begin{equation}
	\mathbf{h}_{\veck}=(\lambda\sin k_x, -\lambda\sin k_y, \mu-2(\cos k_x+\cos k_y))^T.
    	\label{eqn:hvec}
\end{equation}
Due to PBCs, we can consider each quasimomentum mode \(\veck\) as an independent two-level system described by Eqn \eqref{eqn:ham} \cite{dziarmaga}. 

\begin{figure}
    \centering
    \includegraphics[width=0.8\columnwidth]{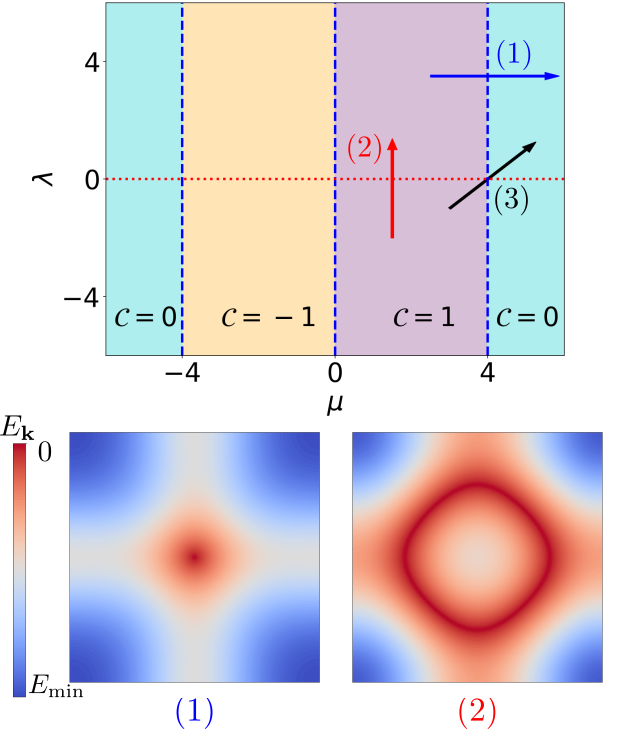}
    \caption{Top panel: The phase diagram for the Hamiltonian given in Eqn (\ref{eqn:ham})  \cite{paolothesis}. The blue dashed and red dotted lines are the phase boundaries for Dirac and nodal loop transitions, respectively. 
    The labelled arrows show the representative quench trajectories across: (1) Dirac transition, (2) nodal loop transition and (3) multicritical point (MCP).
    Bottom panels: The dispersion of the lower band $E_{-,\mathbf{k}}$ given by Eqn (\ref{eqn:disp})  showing (1) Dirac and (2) nodal loop gap closures in red. Here \(E_{\text{min}}={\rm min}(E_{-,\mathbf{k}})\).}
    \label{fig:phasediag}
\end{figure}

This model manifests a rich phase diagram as a function of $\lambda$ and $\mu$, as illustrated in Fig. \ref{fig:phasediag}.
At \(\mu=0, \pm4\), we recover a prototypical Dirac model. In this model, the spectrum, given in Eqn (\ref{eqn:disp}), exhibits a linear gap closure about one or more of the high symmetry points (HSPs) $\veck_0$ obeying $\veck_0=-\veck_0$. 
In this model, the HSPs are situated at \(\veck_0=(0,0),  (0,\pi),(\pi,0)\) and \((\pi,\pi)\) in the Brillouin zone (BZ). 
For \(\lambda=0\), we recover a nodal loop semimetal, where the parameter \(\mu\) controls the size and shape of the nodal loop. 
In what follows, we will examine the criticality and defect dynamics by performing quenches in both $\mu$ and $\lambda$. 
Such quenches are physically realizable in quantum-engineered systems \cite{ucquench, scquench, becquench, latticequench}. 

The topological phases of this model can be characterized by the Chern number \cite{toporev}
\begin{equation}
    	\mathcal{C}=-\frac{1}{2\pi}\int_{BZ}\diff^2\veck \: \mathbf{\hat{h}}_{\veck}\cdot\left(\partial k_x\mathbf{\hat{h}}_{\veck}\times\partial k_y\mathbf{\hat{h}}_{\veck}\right)
    	\label{eqn:chern}
\end{equation}
where \(\mathbf{\hat{h}}_{\veck}\) is a unit vector in the direction of \(\mathbf{h}_{\veck}\) given in Eqn (\ref{eqn:hvec}). 
We remark that, strictly speaking, the Chern number is ill-defined at $\lambda=0$ when the nodal loop gap closure occurs and more general invariants have been proposed to circumvent this problem \cite{ham1}.
Our analysis of quench dynamics across TPTs is however not affected by this issue.
The nature of the gap closures at the TPTs lead to different topological universality classes as classified by CRG \cite{crg}.
This model therefore provides an ideal playground to analyze and probe topological criticality via quenches.

\section{Quench dynamics} 
\label{sec:methods}
In this section, we outline the methodologies we use to study quench dynamics through both linear and sudden quench protocols. 
These help us study both the defect density scaling and unravel the sensitive dependence of defect dynamics and dynamical quantum phase transitions (DQPTs) on the initial conditions.
To study the density of the defects and its Kibble-Zurek (KZ)-like scaling, we consider a linear ramp of the parameters \(\lambda\) and/or \(\mu\) from \(t\rightarrow-\infty\) to \(t\rightarrow\infty\) with quench rate \(\tau^{-1}\). 
The various types of quenches across the TPTs are indicated by arrows in Fig. \ref{fig:phasediag}. 
In the sudden quench protocols, an initial state taken to be the ground state of the initial Hamiltonian \(H_i\) is  time-evolved with a final Hamiltonian \(H_f\) corresponding to a different topological phase.
 
\subsection{Defect density and its scaling}
When a system is linearly quenched across a critical point separating phases with and without spontaneous symmetry breaking, topological defects are generated in the system.
The KZ mechanism \cite{kzm} relates the scaling of the ensuing defect density to the underlying equilibrium   critical exponents characterizing the critical point.
More specifically, as the control parameter \(g\) approaches its critical value \(g_c\), both the correlation length \(\xi\) and the relaxation time \(\tau_{\text{rel}}\) diverge with different critical exponents \cite{aia}, 
\begin{equation}\label{eqn:rel} 
\begin{split}
    \xi &\sim\left|\frac{g-g_c}{g_c}\right|^{-\nu} \\
	\tau_{\text{rel}} &\sim\left|\frac{g-g_c}{g_c}\right|^{-z\nu},
\end{split}
\end{equation}
where $z$ is the dynamical critical exponent.   

For a linear parameter ramp which traverses the critical point, we have  \(\frac{g-g_c}{g_c}\sim\frac{t}{\tau}\). Here, the KZ analysis predicts that 
the density of the bulk  topological defects generated  in a \(d\)-dimensional system  satisfies the following scaling law \cite{dziarmaga}
\begin{equation}
	\ntop\sim\tau^{-\frac{d\nu}{1+z\nu}}
	\label{eqn:kz}
\end{equation}
as the system approaches the gap closure.

In the case of TPTs, though no symmetry-broken order parameter and its associated correlation length exists, the energy gap nonetheless goes to zero at the transition and we still expect defect formation.
For noninteracting systems, this defect density can be estimated in a straightforward manner.
As discussed in Section \ref{sec:model}, the lattice model is just a collection of independent two-level modes in momentum space.
Under a linear quench, the full dynamics reduces to a study of the Landau-Zener transition of each two-level mode \cite{dziarmaga}. 
The most general Landau-Zener probability \(P_\veck\) \cite{zener, aia1} of each \(\veck\) mode is given by the Dykhne formula \cite{dykhne}
\begin{equation}
	P_{\veck}(\tau) \sim\exp\left(-\frac{2}{\hbar}\left|\mathbf{Im}\int_\gamma \diff z~E_\veck(z,\tau)\right|\right)
        \label{eqn:dykhne}
\end{equation}
where we will set \(\hbar=1\) hereafter. 
In Eqn (\ref{eqn:dykhne}), \(E_\veck(z,\tau)\) is the analytically-continued energy dispersion with the complexified time \(t\rightarrow z=z_R+iz_I\), where the $t$ and $\tau$-dependences enter Eqn (\ref{eqn:disp}) through a linear ramp of parameters of the form $\frac{t}{\tau}$. The integration path \(\gamma\) is chosen to describe the time evolution from \(z_R\rightarrow-\infty\) to \(z_R\rightarrow+\infty\) along the real axis. 
In the thermodynamic limit, the density of the bulk excitations generated across the transition is given by  \cite{kz}
\begin{equation}
        \ntop(\tau)\sim\int_{BZ} \diff^2\veck \: P_{\veck}(\tau).
        \label{eqn:ntop}
\end{equation}

Eqn (\ref{eqn:ntop}) is expected to yield  \(\ntop\sim\tau^{-\alpha}\) to leading order in the limit of large $\tau$. 
For our model with \(z\nu=1\) \cite{kyang} and \(d=2\), a comparison to Eqn \eqref{eqn:kz} implies
\begin{equation}
\alpha=\nu
\label{eqn:alpha}
\end{equation}
if the KZ prediction holds.  
This will permit a direct comparison between the exponents obtained from CRG and from the quantum quench. \\

\subsection{Defect dynamics} \label{sec:method_defects} 

Besides extracting information about criticality via the KZ scaling, the TPTs can also be analyzed in terms of defect generation.
We will thus study whether the nature of the gap closure impacts the dynamics of the defects generated when the system is suddenly quenched across TPTs. 
For a linear gap closure, such as in the Haldane model, topological defects manifest as static and dynamical phase vortices in momentum space \cite{defectsexp1, defectsexp2, defectsexp3,defectspra}. 
The static vortices are immobile, while the dynamical vortices move around the Brillouin zone (BZ) with time. 

In the sudden quench protocol, the time-evolved wave function for a $\veck$ mode is given by $|\psi_\veck(t)\rangle=\exp\left(-iH_{f,\veck}t\right)|\psi_{0,\veck}\rangle$, where \(|\psi_{0,\veck}\rangle\) is the initial state. 
Using the  Bloch sphere representation \cite{defectspra}, the wave function can be written as 
\begin{equation}
	|\psi_\veck(t)\rangle=\begin{pmatrix}
        		\sin\left(\frac{\theta_\veck(t)}{2}\right)\\
        		-\cos\left(\frac{\theta_\veck(t)}{2}\right)e^{i\varphi_\veck(t)}
        \end{pmatrix}.
        \label{eqn:bloch} 
\end{equation}
The static and dynamical vortices can be described via the winding of the azimuthal phase $\varphi_{\veck}$.
A dynamical vortex appears when the modes cross one of the poles of the Bloch sphere, where $\varphi_{\veck}$ is ill-defined and a \(2\pi\)-jump in $\varphi_{\veck}$ occurs.

\subsection{Dynamical quantum phase transitions} \label{sec:dqpt}
Defect generation and dynamics  are both intricately linked to the notion of DQPT, which is characterized by  emergent nonanalyticities at critical times \cite{dqpt,defectspra}
 in  the Loschmidt  echo  \cite{dqpt, dqpt3} defined as:
    \begin{equation}
        \mathcal{L}(t)\equiv |\mathcal{G}(t)|^2=|\langle\psi_0|\psi(t)\rangle|^2
        \label{eqn:gk}
    \end{equation}
where $\mathcal{G}(t)$ is the Loschmidt amplitude.

Nonanalyticities manifest as complex zeros, also known as Fisher zeros, of \(\mathcal{G}(z)\) in the complex time domain \(t\rightarrow z=z_R+iz_I\) \cite{dqpt}.
In the thermodynamic limit, Fisher zeros accumulate on lines or areas, and DQPTs are said to occur whenever these lines or areas cross the real-time axis \cite{dqpt}. 
In other words, DQPTs occur whenever \(\mathcal{G}(t)=0\), where \(t\in\mathbb{R}\). 

For a translationally invariant and particle-hole symmetric noninteracting fermionic system, the Loschmidt amplitude factorizes as \(\mathcal{G}(t)=\Pi_{\veck}\mathcal{G}_{\veck}(t)\) \cite{dqpt, dqpt3}.
For the sudden quench protocol with \(\mathbf{h}_{i,\veck}\to \mathbf{h}_{f,\veck}\), the zeros of Eqn (\ref{eqn:gk}) are given by the modes \(\veck_c\) satisfying the conditions \cite{dqpt_condition}
    \begin{equation}
        \mathbf{h}_{i, \veck_c}\cdot\mathbf{h}_{f, \veck_c}=0
        \label{eqn:loszeros}
    \end{equation}
and
\begin{equation}
  |\mathbf{h}_{f, \kc}|=\frac{n\pi}{2t}, n=1,3,5...
    \label{eqn:tdqpt}
\end{equation}
at a given time $t$ \cite{dqpt}.
In combination with Eqn (\ref{eqn:bloch}),  these conditions  correspond to a reflection about the equatorial plane on the Bloch sphere: for a given mode \(\kc\) starting at polar angle  \(\theta_0\), DQPT occurs when its polar angle becomes $\pi-\theta_0$.

Alternatively, DQPTs can be characterized by a \(2\pi\)-jump \cite{dqpt} in the Pancharatnam geometric phase (PGP) \cite{pancharatnam, pgp}. 
 The PGP is defined for each decoupled momentum mode as
    \begin{equation}
        \phi^P_\veck(t)=\text{arg}\left[\gk(t)\right]-\phi^{dyn}_{\veck}(t)
        \label{eqn:pgp}
    \end{equation}
where 
      \(\phi^{dyn}_{\veck}(t)=-\int_0^t \diff s~ \langle\psi_{\veck}(s)|H_{\veck}(s)|\psi_{\veck}(s)\rangle\) is the dynamical phase.  In the following, we will use the procedures outlined in this section to study the quench dynamics of different classes of TPTs.


\section{Dirac transition}\label{sec:dirac}
Quenches across Dirac gap closures have been investigated in previous literature by considering the low-energy theory close to the gap closure point \cite{kyang}. 
For our model featuring both nodal loop and Dirac gap closures, a natural question arising is whether this added complexity influences the quench dynamics. 

\subsection{Scaling of defect density} \label{sec:diracscaling}

The model described by Eqn (\ref{eqn:ham}) features a first-order Dirac transition where the energy gap scales linearly close to  a HSP \(\veck_0\). 
The CRG predicts that \(\nu=\frac{1}{n}\) ~\cite{crg} for $n$-th order Dirac transitions where the energy gap scales as \(\Delta(\veck)\sim|\veck-\veck_0|^n,~n\in\mathbb{Z}^+\). 
Based on the CRG prediction, the Kibble-Zurek (KZ) scaling given by Eqns \eqref{eqn:kz} and \eqref{eqn:alpha} predicts that the defect density \(\ntop\sim\tau^{-1}\) for large quench times $\tau$.

To verify this predicted KZ scaling across the Dirac transition, we first calculate the transition probability using Eqn (\ref{eqn:dykhne}). 
By setting \(\mu=\frac{t}{\tau}\) at fixed \(\lambda\), we obtain
\begin{equation}
       	P_\veck(\tau)=\exp\left(-\pi\tau\lambda^{2}(\sin^2k_x+\sin^2k_y)\right).
        \label{eqn:probdirac}
\end{equation}
Using a saddle point approximation \cite{spa} for Eqn \eqref{eqn:ntop}, valid for \(\pi\tau\lambda^{2} \gg 1\), we obtain 
    \begin{equation}
        \ntop=\frac{A_1}{\pi\tau\lambda^{2}} + \frac{A_2}{(\pi\tau\lambda^{2})^2}+\mathcal{O} \left( \tau^{-3} \right)
        \label{eqn:spadirac1}
    \end{equation}
where the \(\tau\)-dependence of the coefficients  \(A_1\) and \(A_2\) is negligible when \(\pi\tau\lambda^{2} \gg1\).
Therefore, we recover the expected KZ scaling behavior $n_{\text{topo}} \sim \tau^{-1}$ to leading order.
We have verified that this consistency between CRG and KZ predictions is also valid for higher order Dirac gap closures. 
This shows, to the leading order, that the KZ scaling is insensitive to the modes away from the gap closure point.

\subsection{Dynamical phase vortices and DQPTs} \label{sec:defects_dirac} 
The complexity of our model allows for a rich set of initial states to be realized. 
In the sudden quench protocol across the Dirac transition, we take \(\mu_i<0\) in the initial Hamiltonian, such that the modes are spread across the Bloch sphere. 
The modes corresponding to the HSPs always lie on the south pole regardless of $\mu_i$. 
They correspond to the static vortices in our model.

At a given time $t$, the dynamical vortices appear at the modes \(\kst\) satisfying the conditions \(\theta_{\kst}(t)=0\) and \(|\mathbf{h}_{f, \kst}|=\frac{n\pi}{2t}, n=1,3,5...\), which are generalizations of the conditions outlined in Ref.~~\cite{defectspra}. 
By solving these conditions exactly, we find that the trajectory of the dynamical vortices is given by  a shifted nodal loop described by
\begin{equation}
    	\mu\rightarrow\mu_n(t)=\mu_f-\frac{1}{c(\mu_f-\mu_i)}\left(\frac{n\pi}{2t}\right)^2, n=1,3,5...
    	\label{eqn:ripple_dirac}
\end{equation}
where \(c=2\) is a constant and $\mu_f>0$ is the parameter of the final Hamiltonian.

\begin{figure}
    \centering
    \includegraphics[width=0.95\columnwidth]{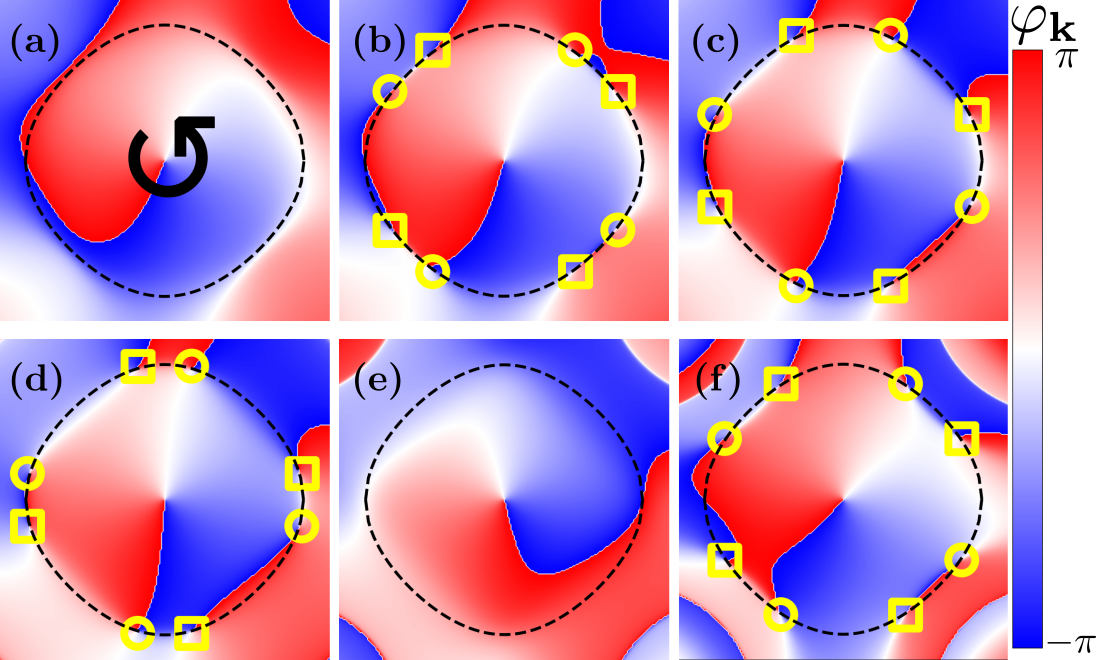}
    \caption{The azimuthal phase \(\varphi_\veck(t)\) for sudden quench with \(\mu_i\ra-\infty\) at \textbf{(a)} \(t=0.6\), \textbf{(b)} \(t=0.65\), \textbf{(c)} \(t=0.7\), \textbf{(d)} \(t=0.75\), \textbf{(e)} \(t=0.9\) and \textbf{(f)} \(t=2.0\).
    The vortices are identified as the winding of \(\varphi_\veck\) from \(-\pi\) to \(\pi\). \textbf{(a)} shows the  winding direction of the static vortex at $\veck=(0,0)$.
    The dynamical vortex-antivortex pairs consist of anti-clockwise and clockwise windings, which are marked by yellow squares and circles, respectively.
    \textbf{(b)-(f)} show the static trajectory of the dynamical vortices along the underlying nodal loop shown by the dashed lines. The generated vortex pairs travel along the static trajectory and annihilate with each other before reappearing.
    The figures show a window of \(k_x,k_y\in[-0.6\pi,0.6\pi]\).}
    \label{fig:dphi}
\end{figure}

\begin{figure}
	\centering
    	\includegraphics[width=0.95\columnwidth]{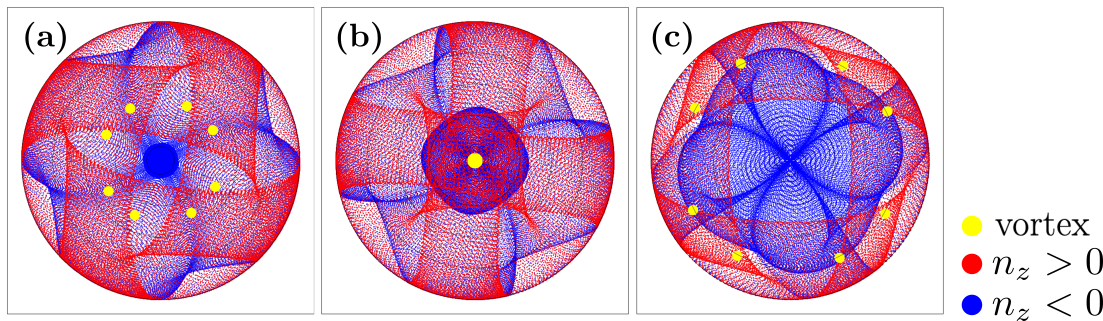}
    	\caption{2D projections of the Bloch sphere on the \(x\text{-}y\) plane for sudden quench starting from $\mu_i\ra-\infty$ at \textbf{(a)} \(t=0.6\), \textbf{(b)} \(t=0.7\) and \textbf{(c)} \(t=0.9\). The corresponding azimuthal phase is shown in Fig. \ref{fig:dphi} \textbf{(a)},  \textbf{(c)} and  \textbf{(e)}, respectively.   
    	Blue and red dots indicate the modes lying in the southern and northern hemispheres, respectively. The yellow dots indicate the modes corresponding to dynamical vortices $\kst$ at \(t=0.7\). 
    	For \(\mu_i\ra-\infty\), the modes start from the south pole at $t=0$ and spread to the northern hemisphere as indicated in \textbf{(a)}. Dynamical vortices are generated when some \(\veck\) modes reach the north pole of the Bloch sphere, and disappear when no mode lies there anymore.
    }
    	\label{fig:dbs}
\end{figure}

In the general case of finite $\mu_i<0$, the trajectory of the defects becomes time-dependent, as shown in Eqn \eqref{eqn:ripple_dirac}, with the shape of the dynamical trajectory being dictated by $\mu_f-\mu_i$.
Furthermore, the trajectories exhibit a ripple-like effect stemming from a discrete shift in the nodal loop characterized by $n$.
This is in stark contrast to the static trajectories studied in the literature \cite{defectspra, defectsexp1, dqpt}.
There, a fully-polarized initial state consisting of all modes lying on the pole of the Bloch sphere was considered. 
In our model, this polarized initial state corresponds to the limit of  $\mu_i \to -\infty$. 
The effects of the underlying nodal loop are palpable even with the polarized initial state, where the static trajectory is given by the original nodal loop with $\mu_f$ \cite{defectsexp1, defectspra}, as shown in Fig. \ref{fig:dphi}. 
The trajectory of the dynamical vortices on the Bloch sphere is shown in Fig. \ref{fig:dbs}.

For the  polarized initial state \cite{defectspra, dqpt}, the  creation of dynamical vortices after a sudden quench is a marker of  DQPTs. For a general initial condition, we find, using Eqns (\ref{eqn:loszeros}) and (\ref{eqn:tdqpt}), that the dynamical trajectory of the zeros $\veck_c$, where $\mathcal{G}_{\veck_c}(t)=0$ for a given time \(t\), is given by Eqn (\ref{eqn:ripple_dirac}), but now with \(c=1\).  
When \(\mu_i\rightarrow-\infty\), we indeed recover \(\mu_n(t)\rightarrow\mu_f\), verifying that in this case the trajectories of the dynamical vortices and DQPTs coincide. 

\begin{figure}
    \centering
    \includegraphics[width=0.8\columnwidth]{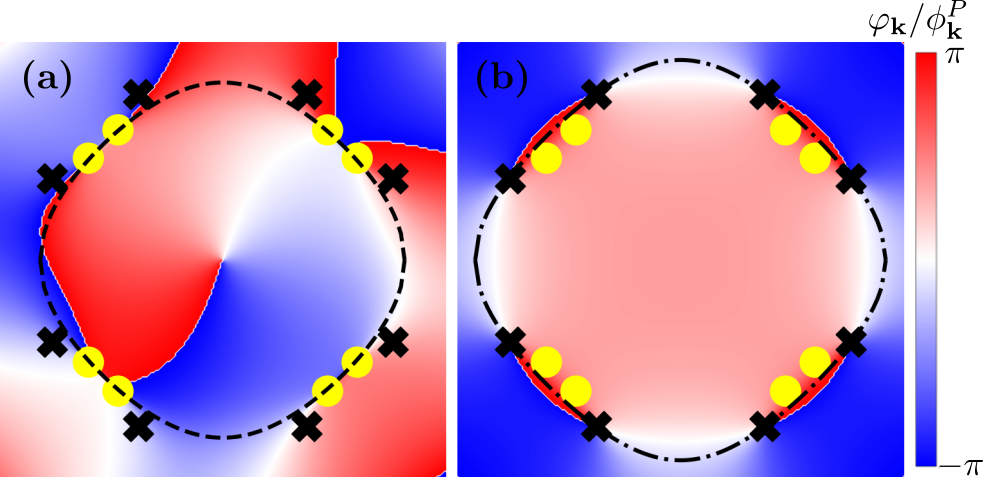}
    \caption{Plots of \textbf{(a)} the azimuthal phase $\varphi_{\mathbf{k}}$, and \textbf{(b)} the Pancharatnam geometric phase $\phi^P_{\veck}$ at $t=0.6$ for a sudden quench starting from $\mu_i=-6$. This initial condition consists of a spread of modes across both hemispheres. 
    The yellow circles indicate the  \(\kst\) modes where dynamical vortices appear.
    The black crosses indicate the $\kc$ modes where the $\phi^P_{\veck}$ shows a \(2\pi\)-jump.
    The shifted nodal loops corresponding to the trajectories of $\kst$ and $\kc$ are given by Eqn (\ref{eqn:ripple_dirac}) with $c=2$ (dashed line in \textbf{(a)}) and $c=1$ (dash-dotted line in \textbf{(b)}), respectively. 
     The figures show a window of \(k_x,k_y\in[-0.7\pi, 0.7\pi]\).
    }
    \label{fig:dqpt}
\end{figure}

As mentioned before, within the Bloch sphere picture, DQPTs are concomitant with a reflection of the modes about the equatorial plane. 
When the initial state is completely polarized and hence all modes lie on the south pole, DQPT occurs for the modes crossing the north pole at a given time. 
Coincidentally, these modes can also be identified as the dynamical vortices, since the azimuthal phase is ill-defined at the north pole. 
For general initial conditions, however, such a correspondence no longer holds.
This is because the modes $\kst$ and $\kc$, where the dynamical vortices and DQPTs respectively occur for a given time $t$, do not coincide in general. 
Only the PGP, shown in Fig. \ref{fig:dqpt}, is the true physical quantity which characterizes the DQPTs through the zeros of the Loschmidt echo.

To summarize, in contrast to the defect density which obeys the usual KZ scaling for Dirac transitions, the defect \emph{dynamics} are indeed sensitive to the underlying nodal loop structure of the Hamiltonian, even in the limit of a polarized initial state.

\section{Nodal loop transition}\label{sec:nl}

We now study the quench dynamics across the extended nodal loop gap closure, as shown in Fig. \ref{fig:phasediag}. 
In this case, the presence of the extended nodal loop will modify both the scaling exponent and the dynamics of the defect generation.

\subsection{Scaling of defect density} \label{sec:scalingnl}
We again consider a linear ramp, this time with \(\lambda=\frac{t}{\tau}\) at fixed \(\mu\) values.
For this quench procedure, the nodal loop transition probability calculated using Eqn (\ref{eqn:dykhne}) is 
    \begin{equation}
    P_\veck(\tau)=\exp\left(-\pi\tau\frac{(\mu-2(\cos k_x+\cos k_y))^2}{\sqrt{\sin^2k_x+\sin^2k_y}}\right).
        \label{eqn:probnl}
    \end{equation}

Two regimes of  defect scaling are now seen:
for \(-4<\mu<4, \mu\neq0\), the saddle point approximation \cite{spa} for the defect density, valid in the limit of \(\pi\tau\gg1\), yields 
    \begin{equation}
        n_{\text{topo}}\approx\frac{C_1}{\sqrt{\pi\tau}}+\frac{C_2}{(\pi\tau)^{3/2}}+\mathcal{O}(\tau)^{-5/2},
        \label{eqn:spanl1}
    \end{equation}
while for \(\mu=\pm4\) we obtain 
        \begin{equation}
        n_{\text{topo}}\approx\frac{D_1}{(\pi\tau)^{2/3}}+\frac{D_2}{(\pi\tau)^2}+\mathcal{O} (\tau)^{-10/3}.
        \label{eqn:spanl2}
    \end{equation}

As expected from KZ scaling (Eqn \eqref{eqn:kz}), we see that \(\ntop\sim\tau^{-\alpha}\) to the leading order in both cases. 
For a truly extended gap closure, which is the first case, we obtain \(\alpha=\frac{1}{2}\). 
At $\mu=\pm4$, instead, the nodal loop collapses to a point at the HSPs where the Dirac and nodal loop gap closures coincide. 
As such, a quench with $\mu=\pm4$ corresponds to a quench along the phase boundary. 
We obtain \(\alpha=\frac{2}{3}\) for such a gap closure.
This will be discussed further in Section \ref{sec:mcp}, where we investigate multicriticality.
Contrary to the Dirac transition, the interpretation of the critical exponents \(z\) and \(\nu\) is not clear in these cases.

In analogy with higher order Dirac gap closures \(\Delta(\veck)\sim|\veck-\veck_0|^n\) about a point $\veck_0$, we can also generalize the nodal loop gap closure to \(n\)-th order.
This is best done in polar coordinates, \(\Delta(\theta)\sim|r-r_0(\theta)|^n\), where \(r_0(\theta)\) is the angle-dependent nodal loop radius. 
Such a closure can be achieved by the replacement  \(\mu-\xi_{\veck}\ra(\mu-\xi_{\veck})^n\) in Eqn (\ref{eqn:ham}). 
We find that the defect density then scales as \(\ntop\sim\tau^{-1/2n}\).
Our result is consistent with the expression given in Ref.~\cite{dziarmaga}, $n_{\text{topo}}\sim\tau^{-\frac{d-m}{z_\Delta}}$, where \(d\) is the physical dimension, \(m\) is the dimension of the Fermi manifold, and \(z_\Delta=2n\) for the \(n\)-th order nodal loop. 
Comparing with the Dirac case where the exponent is $1/n$, we see that an extended gap closure increases the defect density to the leading order, in the limit of large $\tau$.
This can have important consequences for the behavior of nodal loop semimetals and their edge states.

The factor of $\frac{1}{2}$ in the scaling can be interpreted as the modification \(d\rightarrow d-m\) in the original KZ scaling, where \(m=1\) is the dimension of the gap closure \cite{dziarmaga}. 
This reduction of the effective dimension is a direct consequence of the extended nature of the gap closure, since the energy gap scales similarly for both Dirac and nodal loop transitions at the local level with fixed $\theta$.
The exponent $\alpha=1/2$ is also consistent with some CRG results obtained for a periodically driven Chern insulator \cite{nlfloquet}, where a nodal loop gap closure appears in the effective Floquet Hamiltonian due to emergent symmetries.
There, the curvature function was found to scale with an exponent $\tilde{\nu}_x = 1/2$ along the nodal loop, and with $\nu_y=1$ perpendicular to it. 
Our study then reveals that the CRG scaling of the curvature function matches with the KZ defect generation only in azimuthal direction, suggesting that defects in that Floquet problem should be preferentially generated along the nodal loop and not across it.

\subsection{Dynamical phase vortices and DQPTs} \label{sec:defectsnl}
To investigate the defect  dynamics across nodal loop transitions, we choose \(\lambda_i<0\) which corresponds to a general initial condition discussed in Sec. \ref{sec:defects_dirac}.

\begin{figure}
    	\centering
    	\includegraphics[width=0.95\columnwidth]{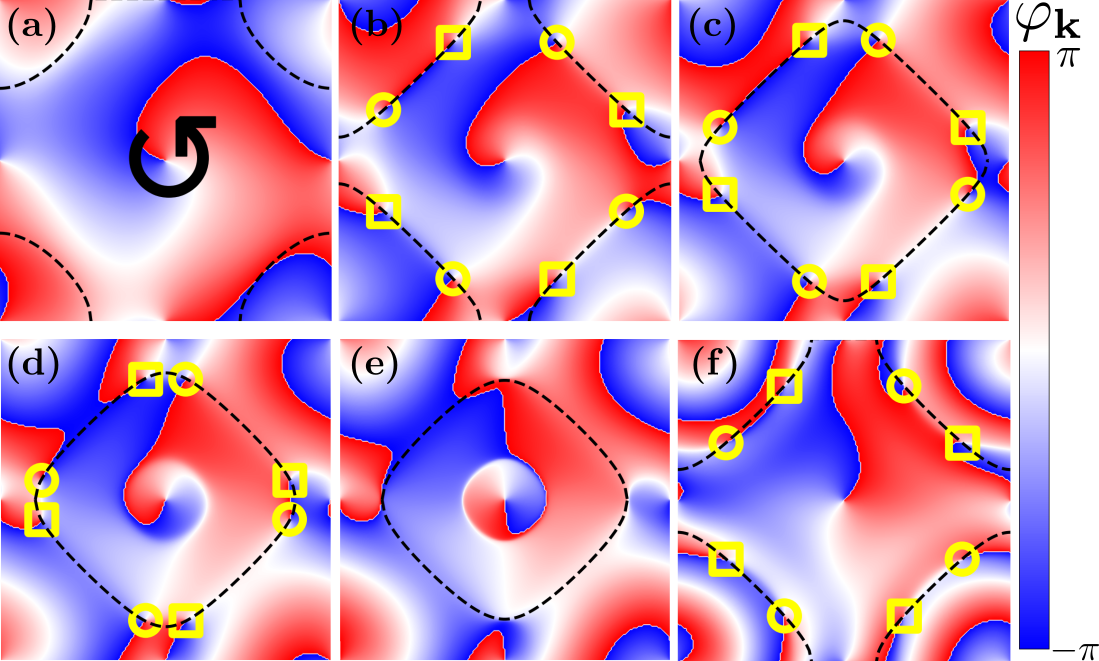}
    	   \caption{The azimuthal phase \(\varphi_\veck(t)\) for sudden quench with \(\lambda_i\ra-\infty\) at \textbf{(a)} \(t=0.3\), \textbf{(b)} \(t=0.5\), \textbf{(c)} \(t=0.6\), \textbf{(d)} \(t=0.7\), \textbf{(e)} \(t=0.8\) and \textbf{(f)} \(t=1.4\).
    \textbf{(a)} shows the  winding direction of the static vortex at $\veck=(0,0)$.
   \textbf{(b)-(f)} show the dynamical vortices where the anti-vortices and vortices are marked by yellow squares and circles, respectively.
    They travel along the time-dependent, discretized trajectory given by Eqn \eqref{eqn:traj} with $c=2$, shown by dashed lines. The generated vortex pairs travel along the trajectory given by \(n=1\) until they annihilate with each other (\textbf{(b)-(e)}). They then reappear along the trajectory given by \(n=3\) (\textbf{(f)}).
    The full Brillouin zone, \(k_x,k_y\in[-\pi,\pi]\), is shown in the plots.
  }
    	\label{fig:nphi}
\end{figure}

Much like the Dirac transition, we can identify the HSPs as the static vortices. 
The dynamical vortices again follow a time-dependent trajectory found from the conditions \(\theta_{\kst}(t)=\pi\) and \(|\mathbf{h}_{f, \kst}|=\frac{n\pi}{2t}, n=1,3,5...\). 
This gives
\begin{equation}
    	\mu\ra\mu_n(t)=\mu-\left(\frac{n\pi}{2t}\right)\frac{1}{\sqrt{c\left(1+\frac{\lambda_f}{|\lambda_i|}\right)}}, ~~n=1,3,5...
   	\label{eqn:traj}
\end{equation}
where \(c=2\) is a constant and $\lambda_f>0$ is the parameter of the final Hamiltonian. 
Notice the additional square root contribution, reminiscent of the square-root scaling law found earlier in the linear quench of the nodal loop gap closure.
Again, we observe a ripple-like effect in the time-dependent trajectory characterized by discrete $n$ values.
Here, however, even the limit \(\lambda_i\rightarrow-\infty\) does not give rise to a fully-polarized initial state, and thus manifests time-dependent trajectories of the dynamical defects as shown in Fig.~\ref{fig:nphi}. 
This is an important distinction from the Dirac case which permits a polarized initial state and hence a static defect trajectory. 

\begin{figure}
    	\centering
    	\includegraphics[width=0.8\columnwidth]{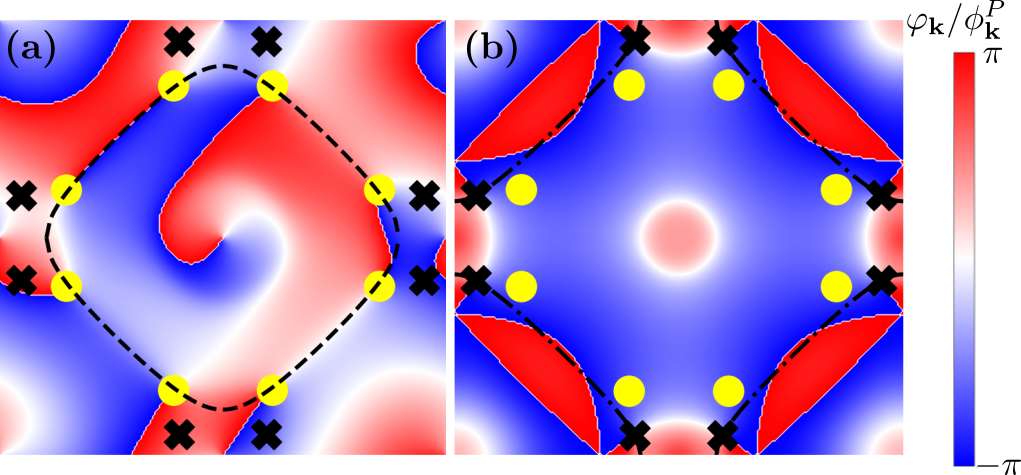}
 \caption{Plots of \textbf{(a)} the azimuthal phase $\varphi_{\mathbf{k}}$, and \textbf{(b)} the Pancharatnam geometric phase $\phi^P_{\veck}$ at $t=0.6$ for a sudden quench starting from $\lambda_i=-6$. 
This initial condition consists of a spread of modes across both hemispheres. 
The yellow circles and the black crosses indicate the \(\kst\) and $\kc$ modes, where dynamical vortices appear and where $\phi^P_{\veck}$ shows a \(2\pi\)-jump, respectively.
    The shifted nodal loops corresponding to the trajectories of $\kst$ and $\kc$ are given by Eqn (\ref{eqn:traj}) with $c=2$ (dashed line in \textbf{(a)}) and $c=1$ (dash-dotted line in \textbf{(b)}), respectively. 
The figures show the full BZ, \(k_x,k_y\in[-\pi, \pi]\).
    }
    	\label{fig:dqpt_nl}
\end{figure}

The zeros $\veck_c$ where $\mathcal{G}_{\veck_c}(t)=0$, found from Eqns and \eqref{eqn:loszeros} and \eqref{eqn:tdqpt}, are  given by Eqn (\ref{eqn:traj}) with \(c=1\). 
This feature is clearly illustrated in Fig.~\ref{fig:dqpt_nl}, which shows that the modes $\kst$ and $\kc$ do not coincide, as in the Dirac case for a general initial condition.
In particular, the zeros are completely wiped out in the limit \(\lambda_i\rightarrow-\infty\). 

\begin{figure}
    \centering
    \includegraphics[width=0.95\columnwidth]{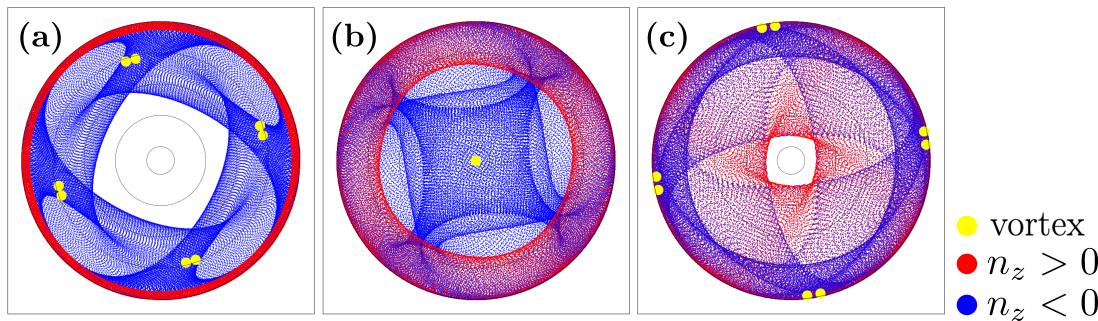}
    \caption{2D projections of the Bloch sphere on the \(x\text{-}y\) plane for sudden quench starting from $\lambda_i\ra-\infty$ at \textbf{(a)} \(t=0.3\), \textbf{(b)} \(t=0.5\) and \textbf{(c)} \(t=0.8\). The corresponding azimuthal phase is shown in Fig. \ref{fig:nphi} \textbf{(a)},  \textbf{(b)} and  \textbf{(e)}, respectively.   
Blue and red dots indicate the modes lying in the southern and northern hemispheres, respectively. 
The yellow dots indicate the modes corresponding to dynamical vortices $\kst$ at \(t=0.5\). 
Under this initial condition, the modes start from the equator and spread to both hemispheres.
Dynamical vortices are generated when some \(\veck\) modes reach the south pole of the Bloch sphere, and disappear when no mode lies there anymore.
    }
    \label{fig:nbs}
\end{figure}

We see that nodal loop transitions typically lead to complex time-dependent trajectories for the defects, in contrast to what is known in the present literature \cite{defectsexp1, defectspra}.
Consequently, the creation of dynamical vortices at a given time does not imply the occurrence of DQPTs at that mode, since the modes where the dynamical vortices and the zeros of $\gk(t)$ appear do not coincide in general.

\section{Multicriticality}\label{sec:mcp}
Our model  hosts multicritical points (MCPs) at \((\mu,\lambda)=(0,0)\) and \((\mu,\lambda)=(\pm4,0)\) where both Dirac and nodal loop gap closures coexist.
Quench dynamics through MCPs~\cite{mcp1, mcp3, kumarmcp1, kumarmcp2} and along critical surfaces~\cite{mcp4} have been investigated in 1D spin systems and lattice chains, where it has been shown that the scaling of the defects is modified by multicriticality.  
In Sec. \ref{sec:scalingnl}, we have also seen that we obtain different results for a vertical quench along the phase boundary compared to a purely nodal loop transition. 
Here, we address the general case of a diagonal quench through the MCPs \((\mu,\lambda)=(\mu_c,0)\), where $\mu_c=0,~4$. We do so by setting \(\lambda=m(\frac{t}{\tau}-\mu_c)\) and \(\mu=\frac{t}{\tau}\), where \(m\) is the gradient of the quench path in the parameter space.

The corresponding LZ probability is found to be
\begin{align}
        P_{\mathbf{k}}(\tau) &= \exp \Bigg(
        -\pi\tau m^2 \left( \sin^2 k_x+\sin^2k_y \right) \nonumber \\
        & \qquad \quad \times \left( 2(\cos k_x+\cos k_y)-\mu_c\right)^2 \nonumber \\
        & \qquad \quad \times \left(1+m^2(\sin^2 k_x+\sin^2k_y) \right)^{-3/2}
        \Bigg),
	\label{eqn:probdiag}
\end{align}
from which we obtain $n_{\text{topo}}$ in the limit \(\tau m^2\gg1\).
The case of $\mu_c=4$, where the nodal loop collapses to a point, yields the following KZ scaling
\begin{equation}
        n_{\text{topo}}=\frac{I_1}{(\pi\tau m^2)^{1/3}}+\frac{I_2}{(\pi\tau m^2)^{2/3}}+\mathcal{O}\left(\pi\tau m^2 \right)^{-1}
        \label{eqn:spadiag1}
\end{equation}
where the coefficients \(I_1\sim1\) and \(I_2\sim m^2\) have a negligible \(\tau\)-dependence. 
We obtain different scaling  regimes for $\ntop$ depending on the value of $m$. 
For $m\sim1$,  since the magnitudes of \(I_1\) and \(I_2\) are comparable, to leading order \(\ntop\sim\tau^{-1/3}\). For \(m\gg1\), the second term in Eqn (\ref{eqn:spadiag1}) starts to dominate and we  observe a change in the leading order scaling to \(\ntop\sim\tau^{-2/3}\), which as expected is the power law behavior for a critical quench as in Eqn (\ref{eqn:spanl2}). 

For the case of $\mu_c=0$, we obtain 
\begin{equation}
        n_{\text{topo}}=\frac{I_1}{(\pi\tau m^2)^{1/3}}+\frac{I_2}{(\pi\tau m^2)^{1/2}}+\mathcal{O}\left(\pi\tau m^2 \right)^{-2/3}
        \label{eqn:spadiag2}
\end{equation}
where the leading and next order contributions are due to Dirac and nodal loop gap closures, respectively. The coefficients scale as $I_1\sim1$ and $I_2\sim m^{3/2}$, which again lead to a path-dependent behavior. For $m\sim1$, we observe $n_{\text{topo}}\sim\tau^{-1/3}$ due to the Dirac gap closure, but at large $m\gg1$, the power law is dominated by $n_{\text{topo}}\sim\tau^{-1/2}$ due to the nodal loop gap closure.

The condition $\tau m^2\gg1$ can be interpreted as the scaling of the effective quench timescale by \(m^2\), so the gradient directly affects the regime of \(\tau\) where we can access the KZ scaling. 
This is reminiscent of  the rescaling  \(\tau \to \tau\lambda^{2}\)  seen for Dirac transitions.
We expect the defect dynamics  to be qualitatively similar  to the Dirac and the nodal loop case, where dynamical phase vortices appear whenever a pole of the Bloch sphere is reached during the evolution of the states away from HSPs.

\section{Conclusions and Outlook}
\label{sec:outlook}
In this work, we have analyzed the defect scaling and dynamics across various topological phase transitions (TPTs)  in a model hosting different universality classes.

 For  TPTs  characterized by Dirac-like gap closures, we recover the expected   Kibble-Zurek (KZ) scaling of the defect density with exponents consistent with those predicted by the curvature renormalization group (CRG) \cite{crg}. 
On the other hand, the nodal loop transitions feature a reduced  KZ scaling exponent stemming from the extended nature of the gap closure, and compatible with the scaling of the curvature function seen in related Floquet models \cite{nlfloquet}.
Quenches across multicritical points reveal path-dependent behaviors of the scaling exponents depending on the gradient of the path in parameter space. 

Nonetheless,  under a sudden quench, the  complex nature of the underlying band dispersion encompassing a nodal loop manifests via a sensitive dependence of the defect dynamics on the initial condition.  In general, we find that the modes where the dynamical defects appear do not coincide with those where DQPTs occur, with the exception of a fully-polarized initial state discussed in the bulk of the present literature \cite{defectsexp1, defectsexp2, defectsexp3, defectspra, dqpt}. The nodal loop, even when much higher in energy than the Dirac gap, is responsible for a  surprising  ripple-like effect in the dynamical trajectories of the vortices corresponding to shifted nodal loops. This is again in contrast to the static trajectory in the case of polarized initial conditions. To visualize the defect dynamics, we track the evolution of the states on the Bloch sphere across a sudden quench, which is directly measureable in experiments by Bloch state tomography \cite{defectsexp1, defectsexp2, defectsexp3}. 

To summarize, though the KZ-like scaling of the defect density is still  principally defined by the low energy theory, the correct  time-dependent behavior of the defects requires one to go beyond it.
The work here concerns purely the bulk topology as we consider a system with periodic boundary conditions.
A straightforward extension would be to explore the role of edge states in a system with open boundary conditions \cite{kyang}.
This would be particularly interesting in the case of more exotic topological states seen, for instance, in higher-order topological insulators \cite{hoti1, hoti2}.
Our approach could also be applied to systems out of equilibrium or in contact with an environment. 
A natural question to ask is whether dissipation or thermal effects would impact scaling laws and defect dynamics.
This could be studied in topological non-Hermitian models encoding effective processes of gains and losses.
Alternatively, the Loschmidt echo used in the present study could be replaced by a fidelity measure constructed from the density matrix, and DQPTs and defect generation could be probed accordingly.
Finally, it has been recently unveiled, both theoretically \cite{fernandokz2, fernandokz4} and experimentally \cite{fernandokz1, fernandokz3}, that the full counting statistics of the defects exhibit universal behavior. This signature of universality beyond the average defect number calls for an extension to physics beyond the KZ mechanism in topological systems, which could be studied through the defect dynamics presented in this work.

\acknowledgments

This work is partially funded by EPSRC Grants No. EP/P009565/1.
The authors would like to thank Wei Chen and Yuto Shibata for fruitful discussions. 

\bibliography{quench}

\end{document}